\documentclass[english,twoside,a4paper,10pt]{article}

\usepackage[latin1]{inputenc}
\usepackage[T1]{fontenc}
\usepackage{amsmath}
\usepackage{amsfonts}
\usepackage{graphicx}
\usepackage{subfigure}
\usepackage{a4wide}
\usepackage{amssymb}
\usepackage{fancyhdr}
\usepackage{mathrsfs}
\usepackage[toc,page]{appendix}

\begin{document}

\title{\bf $k$-essence non-minimally coupled with Gauss-Bonnet invariant for inflation}
\author{ 
Ratbay Myrzakulov\footnote{Email: rmyrzakulov@gmail.com},\,\,\,
Lorenzo Sebastiani\footnote{E-mail: lorenzo.sebastiani@unitn.it
}\\
\\
\begin{small}
Department of General \& Theoretical Physics and Eurasian Center for
\end{small}\\
\begin{small} 
Theoretical Physics, Eurasian National University, Astana 010008, Kazakhstan
\end{small}\\
}

\date{}

\maketitle


\begin{abstract}
In this paper, we investigated inflationary solutions for a subclass of Horndeski models where a scalar field is non-minimally coupled with the Gauss-Bonnet invariant. Examples of canonical scalar field and $k$-essence to support the early-time acceleration are considered. The formalism to calculate the perturbations in FRW universe and to derive the spectral index and the tensor-to-scalar ratio is furnished.
\end{abstract}



\tableofcontents
\section{Introduction}

Tracing back the history of our expanding universe, we find an era, after the Big Bang, when temperature and effective energy density are extremelly large, such that we expect that some curvature corrections or some extra-field departing from standard matter and radiation (maybe induced by quantum effects or motivated by string theory) appear in the framework of General Relativity. The evidences of this fact are substantiated by the well-accepted idea according to which, near to the Planck scale, the universe underwent a  short period of strong accelerated expansion, namely the early-time inflation~\cite{Linde, revinflazione}, whose dynamics cannot be described by standard cosmology. Thus,
it is believed that inflation may be due to several factors, the mainly being modifications of the theory of gravitation itself (see Refs.~\cite{R1, R2, R3, R4, R5} for some reviews) and the existence of a scalar field, the inflaton, supporting acceleration~\cite{chaotic, buca1, buca2, buca3, buca4}. 

The Lagrangian of modified gravity is much more involved with respect to the one of General Relativity, and leads to fourth order differential equations,
but in 1974 Horndeski~\cite{Horn} found the most general class of scalar tensor theories (where a scalar field is coupled with gravity) which possesses standard Lagrangian like in the theory of Einstein. Horndeski gravity is quite popular and has been considered in many works, especially in the context of inflationary cosmology~\cite{Def, DeFelice, Kob, Kob2, Qiu, Maselli, EugeniaH, mioH, cognoH, add1, add2, add3, add4}. Moreover, one interesting subclass of Horndeski gravity is represented by the non-minimal coupling of the field with the Gauss-Bonnet four dimensional topological invariant~\cite{GB1, GB2, GB3, GB4, GB5, GBOd}, since the Gauss-Bonnet invariant is strictly connected with the string theory and the trace anomaly and may play an important role in the early-time expansion of our universe.

In the present work, we will consider a class of models where a scalar field supporting inflation is coupled with the Gauss-Bonnet invariant. The theory can be inferred as a special case of Horndeski gravity. At first we will investigate simple models where the inflaton is identified with a canonical scalar field, and therefore we will extend the analysis to some $k$-essence models whose Lagrangian contains non-standard higher order kinetic term~\cite{kess1, kess2, kess3}. 

The paper is organized in the following way. In Section {\bf 2}, we derive our model from the Horndeski Lagrangian and we present the field equations on Friedmann-Robertson-Walker space-time. In Section {\bf 3}, some inflationary solutions are reconstructed for simple scalar field and $k$-essence non minimally coupled with the Gauss-Bonnet invariant.  Section {\bf 4} is devoted to the analysis of the cosmological perturbations during inflation. Here, the spectral index and the tensor-to-scalar ratio are calculated. Conclusions are given in Section {\bf 6}.

We use units of $k_{\mathrm{B}} = c = \hbar = 1$ and 
$8\pi/M_{Pl}^2=1$, where $M_{Pl}$ is the Planck Mass.

\section{The model}

Let us start by considering the following gravitational model,
\begin{equation}
I=\int_\mathcal M dx^4\sqrt{-g}\left[\frac{R}{2}+p(\phi, X)+\xi(\phi) \mathcal G\right]\,,
\quad X=-\frac{g^{\mu\nu}\partial_\mu \phi\partial_\nu\phi}{2}\,,
\label{action}
\end{equation}
where $\mathcal M$ is the space-time manifold, $g$ is the determinant of the metric tensor $g_{\mu\nu}$, $R$ is the Ricci scalar of the Hilbert-Einstein action of General Relativity (GR), $p(\phi, X)$ is a function of the scalar field $\phi$ and its kinetic energy $X$, $\xi(\phi)$ is a function of the field only and $\mathcal G$ is the Gauss-Bonnet four dimensional topological invariant, namely
\begin{equation}
\mathcal G=R^2-4R_{\mu\nu}R^{\mu\nu}+R_{\mu\nu\sigma\xi}R^{\mu\nu\sigma\xi}\,,
\end{equation}
$R_{\mu\nu}$ and $R_{\mu\nu\sigma\xi}$ being the Ricci tensor and the Riemann tensor, respectively.

The Lagrangian in (\ref{action}) represents a special case of Horndeski gravity.
Horndeski models~\cite{Horn} are the most general scalar-tensor theories with field equations at the second order (like in GR) and assume the general form (in vacuum),
\begin{equation}
I=\int_\mathcal M dx^4\sqrt{-g}\left[\frac{R}{2}+\mathcal L_H\right]\,,\quad \mathcal{L}_H=\sum_{i=2}^5\mathcal{L}_i\,,\label{action01}
\end{equation}
where $\mathcal L_H$ collects the Lagrangian of the scalar field $\phi$ and higher order corrections to GR coupled with the field itself,
\begin{equation}
\mathcal{L}_2=P(\phi,X)\,,\nonumber
\end{equation}
\begin{equation}
\mathcal{L}_3=-G_3(\phi,X)\Box\phi\,,\nonumber
\end{equation}
\begin{equation}
\mathcal{L}_4=G_4(\phi,X)R+G_{4,X}[(\Box\phi)^2-(\nabla_\mu \nabla_\nu \phi)(\nabla^\mu \nabla^\nu \phi)]\,,\nonumber
\end{equation}
\begin{eqnarray}
\mathcal{L}_5&=&G_5(\phi,X)G_{\mu\nu}(\nabla^\mu \nabla^\nu \phi)-\frac{1}{6}G_{5,X}[(\Box\phi)^3-\nonumber\\&&
3(\Box\phi)(\nabla_\mu \nabla_\nu \phi)(\nabla^\mu \nabla^\nu \phi)+2(\nabla^\mu \nabla_\alpha\phi)(\nabla^\alpha \nabla_\beta \phi)(\nabla^\beta \nabla_\mu \phi)]\,.
\label{cGR}
\end{eqnarray}
Here,
$G_{\mu\nu}:=R_{\mu\nu}-R g_{\mu\nu}/2$ is the usual Einstein's tensor, while
$P(\phi, X)$ and $G_i(\phi, X)$ with $i=3,4,5$ are functions of the scalar field $\phi$ and its kinetic energy $X$.
If we set~\cite{Kob}
\begin{eqnarray}
P(\phi, X)&=& p(\phi, X)+ 8\frac{d^4\xi(\phi)}{d\phi^4}X^2(3-\log X)\,,\nonumber\\
G_3(\phi, X)&=& 4\frac{d^3\xi(\phi)}{d\phi^3}X(7-3\log X)\,,\nonumber\\
G_4(\phi, X) &=&4\frac{d^2\xi(\phi)}{d\phi^2}X(2-\log X)\,,\nonumber\\
G_5(\phi, X)&=&-4\frac{d\xi(\phi)}{d\phi} \log X\,,
\end{eqnarray}
after integration by part it is possible to recover (\ref{action}). By making the simple choice $p(\phi, X)=X$, we find Einstein-dilaton-Gauss-Bonnet (EdGB) gravity. We also note that if $\xi(\phi)=\text{const}$, the contribution of the Gauss-Bonnet disappears.

In general, $\phi$ can be identified with a $k$-essence field whose stress energy-tensor reads~\cite{kess1, kess2},
\begin{equation}
T^{\mu}_{(\phi)\nu}=(\rho(\phi, X)+p(\phi, X))u^{\mu}u_{\nu}+p(\phi, X)\delta^{\mu}_\nu\,,\quad u_\nu=\frac{\partial_\nu\phi}{\sqrt{2X}}\,,\label{Tfluid}
\end{equation}
such that $p(\phi, X)$ is the effective pressure of $k$-essence and $\rho(\phi, X)$ its energy density, 
\begin{equation}
\rho(\phi, X)=2X\frac{\partial p(\phi, X)}{\partial X}-p(\phi, X)\,.\label{rhophi}
\end{equation}
Canonical scalar fields correspond to $p(\phi, X)=X-V(\phi)$, $V(\phi)$ being a function of the field only, but the Lagrangian of $k$-essence admits a higher order kinetic term.

We will work with flat Friedmann-Robertson-Walker (FRW) metric,
\begin{equation}
ds^2=-dt^2+a(t)^2 d{\bf x}^2\,,\label{metric}
\end{equation}
where $a\equiv a(t)$ is the scale factor depending on the cosmological time. Thus,
\begin{equation}
X=\frac{\dot\phi^2}{2}\,,
\end{equation}
and the Equations of motion (EOMs) are given by~\cite{DeFelice, Kob},
\begin{eqnarray}
3H^2=\rho(\phi, X)
-24H^3\dot\phi\frac{d\xi(\phi)}{d\phi}\,,
\label{EOM1}
\end{eqnarray}
\begin{eqnarray}
-(2\dot H+3H^2)&=&p(\phi, X)
+8H^2\frac{d^2\xi(\phi)}{d\phi^2}\dot\phi^2
+8\frac{d\xi(\phi)}{d\phi}(2H^3\dot\phi+2H\dot H \dot\phi+H^2\ddot\phi)
\,,\label{EOM2}
\end{eqnarray}
where
the dot is the derivative with respect to the time.
The continuity equation of $k$-essence follows from the EOMs and reads
\begin{eqnarray}
&&
\dot\rho(\phi, X)+3H(\rho(\phi,X)+p(\phi, X))=
24\frac{d\xi(\phi)}{d\phi}\dot\phi H^2(\dot H+H^2)\,.
\label{conslaw}
\end{eqnarray}
We note that $\rho(\phi, X)+p(\phi, X)=2X p_X(\phi, X)$ and, on FRW space-time, $\mathcal G=24H^2(H^2+\dot H)$.
If we introduce the $e$-folds number respect to a given time $t_0$, namely
\begin{equation}
N=\log\left[\frac{a(t_0)}{a(t)}\right]\,,\label{N}
\end{equation}
and by taking into account that $dN=-Hdt$, we have from (\ref{EOM1}, \ref{conslaw}),
\begin{equation}
3H^2=\rho(\phi, X)+
24H^4\phi'\frac{d\xi(\phi)}{d\phi}
\,,\label{EOM1bis}
\end{equation}
\begin{equation}
-\rho'(\phi, X)+3H^2\phi'^2(p_X(\phi, X))
=24\frac{d\xi(\phi)}{d\phi}\phi' H^3(H'-H)
\,,\label{conslawbis}
\end{equation}
where the prime index denotes the derivative with respect to $N$ and $X=H^2\phi'^2/2$. The $e$-folds is an useful parameter in inflationary cosmology and, if we identify $t_0$ with the time when acceleration ends, it measures the rate of expansion of the universe at different times $t<t_\text{i}$.

\section{Inflation}

The early-time acceleration is realized in a (quasi)-de Sitter space-time when the Hubble parameter is almost a constant and the field moves slowly. In such a case
Eqs.(\ref{EOM1bis}, \ref{conslawbis}) read, in the slow-roll approximation with $|H'/H|\ll 1$,
$|\phi'|\ll 1$ and $|\phi''|\ll  |\phi'|$,
\begin{equation}
3H^2\simeq \rho(\phi, X)
\,,\quad
\rho'(\phi, X)-3H^2 \phi'^2 p_X(\phi, X)\simeq 
24\frac{d\xi(\phi)}{d\phi}\phi' H^4
\,.\label{eqslowroll}
\end{equation}
By taking $t_\text{0}$ in (\ref{N}) as the time at the end of inflation and by assuming $0\ll N$ when accelerated expansion starts, the behavior of the field can be depicted in the following way: when $0\ll N$, $\phi\ll 0$, while at $N=0$, $\phi\simeq 0^-$, such that $\phi'<0$.
Since the Hubble parameter decreases and tends to zero when $N$ approaches to zero, 
$0<\rho'(\phi, X)$, while $X'<0$ to permit a graceful exit from inflation.

We can also introduce the slow-roll parameter
\begin{equation}
\epsilon=\frac{H'}{H}\,,\label{epsilon}
\end{equation}
which is positive and small during the accelerated phase and is of the order of the unit when acceleration finish leading to a total  $e$-folds 
$\mathcal{N}\equiv N(a(t_\text{i}))$, $t_\text{i}$ being the initial time of inflation, large enough to explain the thermalization of the observable universe. To be specific, it must be $55<\mathcal N<65$. 

Let us consider some examples of simple models reproducing inflation in the considered framework.

\subsection{Canonical scalar field}

In this subsection, we will identify $k$-essence with a canonical scalar field by posing (see also Ref.~\cite{Maselli}),
\begin{equation}
p(\phi, X)=X-V(\phi)\,,\quad \rho(\phi, X)=X+V(\phi)\,,\label{dilaton}
\end{equation}
where $V(\phi)$ is a function of the scalar field only and the kinetic term is standard. During inflation, for large and negative values of the field, it must be
\begin{equation}
X\ll V(\phi\rightarrow-\infty)\,,
\end{equation} 
while at the end, when the field tends to vanish,
\begin{equation}
V(\phi\rightarrow 0^-)\ll X\,.
\end{equation}
At first we will consider the following form of the Hubble parameter,
\begin{equation}
H^2=H_0^2(N+1)\,,\quad \epsilon\simeq\frac{1}{2(N+1)}\,,\label{H1}
\end{equation}
with $H_0$ the value of the Hubble parameter at the end of inflation. We easily see that the $\epsilon$ slow-roll parameter is small when $1\ll N$, namely $H$ is almost a constant. By taking into account that $-dN/dt=H$, in terms of the cosmological time the solution above corresponds to
\begin{equation}
H^2=\frac{H_0^4}{4}(t_0-t)^2\,,
\end{equation}
where $t_0$ is the total time of inflation and $t=0$ at the beginning of the accelerated phase.
Thus, by making use of the equations in (\ref{eqslowroll}) with the slow-roll approximation $|H'/H|\ll1$ and $|\phi''|\ll |\phi'|$, one finds
\begin{equation}
V(\phi)\simeq 3H_0^2(N+1)\,,\quad \phi'\simeq \frac{-24\xi_\phi(\phi)H^4+V_\phi(\phi)}{3H^2}\,.\label{esempio1}
\end{equation}
From the second equation we obtain
\begin{equation}
\phi'^2\simeq\frac{-24\xi'(\phi)H_0^4(N+1)^2+3H_0^2}{3H_0^2(N+1)}\,.
\end{equation}
The non-minimal coupling $\xi(\phi)$ between the field and the Gauss-Bonnet determines the form of the field and therefore the mechanism of the exit from inflation. Since the field must move slowly, a realistic scenario may be given by 
\begin{equation}
\xi'(\phi)=\frac{\xi_0}{(N+1)^{2+\lambda}}\,,\quad
\xi(\phi)=-\frac{1}{(1+\lambda)}\frac{\xi_0}{(N+1)^{1+\lambda}}\,,\quad -1<\lambda\,,\label{24}
\end{equation}
with $|\xi_0|\sim 1/H_0^2$ generic constant\footnote{Here, we remember that we are working with unitary Planck Mass, otherwise $|\xi_0|\sim M_{Pl}^{2}/H_0^2$.} and $\lambda$ a number larger than minus one. As a consequence,
\begin{eqnarray}
\phi&\simeq&\phi_0-\left(\frac{2}{1-\lambda}\right)(N+1)^{\frac{1-\lambda}{2}}\sqrt{-8\xi_0 H_0^2}\,,\quad -1<\lambda<0\,,
\nonumber\\
\phi&\simeq&\phi_0-2\sqrt{1+N}\sqrt{-8\xi_0H_0^2+1}\,,\quad 0=\lambda\,,\nonumber\\
\phi&\simeq&\phi_0-2\sqrt{1+N}\,,\quad 0<\lambda\,,
\end{eqnarray}
where $\phi_0<0$ is the value of the field at the end of the accelerated phase and we require $\xi_0<0$. The explicit reconstruction of the potential leads to
\begin{eqnarray}
V(\phi)&\simeq&3H_0^2\left[\left(\frac{2}{1-\lambda}\right)\sqrt{-8\xi_0 H_0^2}\right]^{\frac{2}{\lambda-1}}(\phi_0-\phi)^{\frac{2}{1-\lambda}}
\,,\quad -1<\lambda<0\,,
\nonumber\\
V(\phi)&\simeq&
3H_0^2 \left[2\sqrt{-8\xi_0H_0^2+1}\right]^{-2}(\phi_0-\phi)^2
\,,\quad 0=\lambda\,,\nonumber\\
V(\phi)&\simeq& \frac{3H_0^2}{4}(\phi-\phi_0)^2\,,\quad 0<\lambda\,,\label{potential1}
\end{eqnarray}
while the relation between the potential and the coupling reads
\begin{equation}
\xi(\phi)=-\frac{\xi_0}{(1+\lambda)}\left[\frac{3H_0^2}{V(\phi)}\right]^{1+\lambda}\,,\label{27}
\end{equation}
and one obtains the explicit form of the Lagrangian.
We see that the contribution of the Gauss-Bonnet significatively enters in the dynamics of inflation only for $-1<\lambda<0$. In this case, $\phi'$ is larger with respect to the classical case without  GR-corrections and the field moves faster. We note that at the end of inflation $\xi(\phi)\simeq-\xi_0/(1+\lambda)\sim 1/H_0^2$, and, when $H<H_0$, the Gauss-Bonnet contribution disappears at the small curvatures of Friedmann universe. 

An other example of viable inflation that we would like to consider is given by

\begin{equation}
H^2=H_\text{dS}^2\left[1-\frac{1}{(N+1)}\right]^2\,,\quad
\epsilon\simeq\frac{1}{(N+1)^2}\,,\label{H2}
\end{equation}
where $H_\text{dS}$ is the (de Sitter) Hubble parameter when $1\ll N$. Here, the relation between the $e$-folds $N$ and the cosmological time is given in an implicit way by
\begin{equation}
N+\log[N]=H_\text{dS}(t_0-t)\,,
\end{equation}
where $t_0$ is the total time of the inflationary expansion.
 Now one has
\begin{equation}
V(\phi)\simeq 3H_\text{dS}^2\left[1-\frac{1}{(N+1)}\right]^2\,,
\end{equation}
and by using the suitable form for the coupling function $\xi(\phi)$,
\begin{equation}
\xi(\phi)=\frac{1}{(1-\lambda)}\frac{\xi_0}{(N+1)^{\lambda-1}}\,,\quad 0<\lambda\,,1\neq\lambda\,,\label{exxi}
\end{equation}
where $|\xi_0|\sim1/H_\text{dS}^2$ is a (negative) constant and $\lambda$ a positive number, we obtain
\begin{equation}
\phi'^2\simeq-\frac{8H_\text{dS}^2\xi_0}{(N+1)^\lambda}+\frac{2}{(N+1)^2}\,.
\end{equation}
The following behaviors of the field are inferred:
\begin{eqnarray}
\phi&\simeq&\phi_0-\left(\frac{2}{2-\lambda}\right)(N+1)^{\frac{2-\lambda}{2}}\sqrt{-8\xi_0 H_{\text{dS}}^2}\,,\quad 0<\lambda<2\,,
\nonumber\\
\phi&\simeq&\phi_0-\sqrt{(-8H_\text{dS}^2\xi_0+2)}\log (N+1)\,,\quad 2=\lambda\,,\nonumber\\
\phi&\simeq&\phi_0-\sqrt{2}\log (N+1)\,,\quad 2<\lambda\,,
\end{eqnarray}
with $\phi_0<0$ the value of the field at the end of inflation again. Thus, the potential may be reconstructed as
\begin{eqnarray}
V(\phi)&=&3H_0^2\left[1-2\left(\frac{2}{(2-\lambda)(\phi_0-\phi)}\sqrt{-8\xi_0 H_\text{dS}^2}\right)^\frac{2}{(2-\lambda)}\right]
\,,\quad 0<\lambda<2\,,
\nonumber\\
V(\phi)&=&
3H_\text{dS}^2\left[1-2\text{e}^{(\phi-\phi_0)/\sqrt{-8H_\text{dS}^2\xi_0+2}}\right]
\,,\quad 2=\lambda\,,\nonumber\\
V(\phi)&=& 
 3H_\text{dS}^2\left[1-2\text{e}^{(\phi-\phi_0)/\sqrt{2}}\right]
\,,\quad 2<\lambda\,,\label{33}
\end{eqnarray}
and the coupling function between the field and the Gauss-Bonnet results to be
\begin{equation}
\xi(\phi)=\frac{2^{1-\lambda}\xi_0}{(1-\lambda)}\left[1-\frac{V(\phi)}{3H_\text{dS}^2}\right]^{\lambda-1}\,.\label{34}
\end{equation}
At the end of inflation, since the potential goes to zero, we have $\xi(\phi)\simeq 2^{1-\lambda}\xi_0/(1-\lambda)\sim \pm 1/H_\text{dS}^2$ and the non-GR effects from the Gauss-Bonnet disappear from the Lagrangian.

\subsection{$k$-essence\label{sub32}}
 
In this subsection, we will extend our study to $k$-essence models, where higher order kinetic term appears in the Lagrangian. At first, we will consider the model whose Lagrangian is of the form,
\begin{equation}
p(\phi, X)=F(X)-V(\phi)\,,\label{35}
\end{equation}
with $F(X)$ and $V(\phi)$ two functions depending on $X$ and $\phi$, separately. The potential $V(\phi)$ supports the de Sitter expansion as long as the magnitude of $\phi$ is almost a constant and the
``kinetic'' part $F(X)$ makes inflation end when $\phi'^2$ increases, such that we must have
\begin{equation}
F(X) \ll V(\phi\rightarrow-\infty)\,,\quad V(\phi\rightarrow 0^-)\ll F(X)\,,\label{limitXV}
\end{equation} 
For example, we may take
\begin{equation}
F(X)=\mu X^\kappa\,,\quad 0<\mu\,, \quad 1\leq\kappa\,,\label{k1}
\end{equation}
with $\kappa\,,\mu$ positive parameters. For $\mu=\kappa=1$ we recover the case of canonical scalar field analyzed in the preceding subsection. 
The effective pressure and energy density of $k$-essence correspond to
\begin{equation}
p(\phi, X)=\mu X^\kappa-V(\phi)\,,\quad \rho(\phi, X)=\mu(2\kappa-1) X^\kappa+V(\phi)\,.\label{k12}
\end{equation}
Thus, in the limit (\ref{limitXV}), from
(\ref{eqslowroll}) we get
\begin{equation}
H^2=\frac{V(\phi)}{3}\,,\quad \phi'^{2\kappa}\simeq\frac{2^{\kappa-1}}{\mu\kappa H^{2(\kappa-1)}}\left[-8\xi'(\phi)H^2+\frac{V'(\phi)}{3H^2}\right]\,.
\end{equation}
By assuming (\ref{24}), where $\lambda$ is still larger than minus one to satisfy the limit (\ref{limitXV}), 
solution in (\ref{H1}) can be reconstructed in the following way
\begin{eqnarray}
\phi&\simeq&\phi_0-\Phi_0\left(\frac{2\kappa}{\kappa-\lambda}\right)(N+1)^{\frac{\kappa-\lambda}{2\kappa}}(-8\xi_0 H_0^2)^{1/(2\kappa)}\,,\quad -1<\lambda<0\,,
\nonumber\\
\phi&\simeq&\phi_0-2\Phi_0\sqrt{1+N}(-8\xi_0H_0^2+1)^{1/(2\kappa)}\,,\quad 0=\lambda\,,\nonumber\\
\phi&\simeq&\phi_0-2\Phi_0\sqrt{1+N}\,,\quad 0<\lambda\,,
\end{eqnarray}
with
\begin{equation}
\Phi_0=\left(\frac{2^{\kappa-1}}{\mu\kappa H_0^{2(\kappa-1)}}\right)^{1/(2\kappa)}\,.
\end{equation}
The corresponding forms of the potential are
\begin{eqnarray}
V(\phi)&\simeq&3H_0^2\left[\Phi_0\left(\frac{2\kappa}{\kappa-\lambda}\right)(-8\xi_0 H_0^2)^{1/(2\kappa)}\right]^{\frac{2\kappa}{\lambda-\kappa}}(\phi_0-\phi)^{\frac{2\kappa}{\kappa-\lambda}}
\,,\quad -1<\lambda<0\,,
\nonumber\\
V(\phi)&\simeq&
3H_0^2 \left[2\Phi_0(-8\xi_0H_0^2+1)^{1/(2\kappa)}\right]^{-2}(\phi_0-\phi)^{2}
\,,\quad 0=\lambda\,,\nonumber\\
V(\phi)&\simeq& \frac{3H_0^2}{4\Phi_0^2}(\phi-\phi_0)^2\,,\quad 0<\lambda\,,\label{43}
\end{eqnarray}
and the coupling between the field and the Gauss-Bonnet results to be as in (\ref{27}).
We observe that, for the considered inflationary solution (\ref{H1}), in the classical framework of GR (or, equivalently, if non-GR corrections are negligible), inflation is realized in the same way by canonical scalar field or $k$-essence in (\ref{k1}). However, if one considers the Gauss-Bonnet contribution (this is the case $-1<\lambda<0$), the dynamics of the field changes leading to a different exit from the early-time acceleration.

In a similar way, for the solution (\ref{H2}) with (\ref{exxi}) we derive ($1<\kappa$),
\begin{eqnarray}
\phi&\simeq&\phi_0-\Phi_0\left(\frac{2\kappa}{2\kappa-\lambda}\right)(N+1)^{\frac{2\kappa-\lambda}{2\kappa}}(-8\xi_0 H_{\text{dS}}^2)^{1/2\kappa}\,,\quad 0<\lambda<2\,,
\nonumber\\
\phi&\simeq&\phi_0-\Phi_0\left(\frac{\kappa}{\kappa-1}\right)(-8\xi_0 H_\text{dS}^2+2)^{1/(2\kappa)}(N+1)^{\frac{\kappa-1}{\kappa}}\,,\quad 2=\lambda\,,\nonumber\\
\phi&\simeq&\phi_0-\Phi_0\left(\frac{\kappa}{\kappa-1}\right)2^{1/(2\kappa)}(N+1)^{\frac{\kappa-1}{\kappa}}\,,\quad 2<\lambda\,,
\end{eqnarray}
where
\begin{equation}
\Phi_0=\left(\frac{2^{\kappa-1}}{\mu\kappa H_\text{dS}^{2(\kappa-1)}}\right)^\frac{1}{2\kappa}\,.\label{46}
\end{equation}
Furthermore, the explicit form of the potential is given by
\begin{eqnarray}
V(\phi)&=&3H_0^2\left[1-2\left(\frac{2\kappa\Phi_0}{(2\kappa-\lambda)(\phi_0-\phi)}(-8\xi_0 H_\text{dS}^2)^{1/(2\kappa)}\right)^\frac{2\kappa}{(2\kappa-\lambda)}\right]
\,,\quad 0<\lambda<2\,,
\nonumber\\
V(\phi)&=&
3H_\text{dS}^2\left[1-2
\left(\frac{\kappa\Phi_0}{(\kappa-1)(\phi_0-\phi)}(-8\xi_0 H_\text{dS}^2+2)^{1/(2\kappa)}\right)^\frac{\kappa}{(\kappa-1)}
\right]
\,,\quad 2=\lambda\,,\nonumber\\
V(\phi)&=& 
3H_\text{dS}^2\left[1-2
\left(\frac{\kappa\Phi_0}{(\kappa-1)(\phi_0-\phi)}(2)^{1/(2\kappa)}\right)^\frac{\kappa}{(\kappa-1)}
\right]
\,,\quad 2<\lambda\,,\label{47}
\end{eqnarray}
and the coupling function between the field and the Gauss-Bonnet reads as in (\ref{34}). In this example we can see that the presence of the higher order kinetic term of $k$-essence modifies inflation with respect to the classical canonical scalar field case even when the contribution of Gauss-Bonnet is negligible.\\
\\
Let us consider now the following form of $k$-essence,
\begin{equation}
p(\phi, X)=g(\phi)X^\kappa-V(\phi)\,,\quad 
\rho(\phi, X)=(2\kappa-1)g(\phi)X^\kappa+V(\phi)\,,\quad1\leq\kappa\,,
\label{k2}
\end{equation}
where $g(\phi)$ and $V(\phi)$ are functions of the field and $\kappa$ is a positive number. When
$g(\phi)=\mu$, we recover (\ref{k1}). 
When
the following limits take place
\begin{equation}
g(\phi\rightarrow\infty)X^\kappa\ll V(\phi\rightarrow\infty)\,,\quad
V(\phi\rightarrow 0^-)\ll g(\phi\rightarrow 0^-)X^\kappa\,,
\end{equation}
equations in (\ref{eqslowroll}) read
\begin{equation}
H^2=\frac{V(\phi)}{3}\,,\quad \phi'^{2\kappa}\simeq\frac{2^{\kappa-1}}{g(\phi)\kappa H^{2(\kappa-1)}}\left[-8\xi'(\phi)H^2+\frac{V'(\phi)}{3H^2}\right]\,.
\end{equation}
To reproduce the accelerated solution (\ref{H1}), we may assume the following Ansatz:
\begin{equation}
g(\phi)=\frac{\mu}{(N+1)^\sigma}
\,,\quad
\xi(\phi)=-\frac{1}{(1+\lambda)}\frac{\xi_0}{(N+1)^{1+\lambda}}\,,\quad -1<\lambda\,,1\neq\lambda\,,0<\sigma<\lambda+\kappa\,,\label{exxi2}
\end{equation}
where $\mu$ is a positive constant, $\xi_0$ a (negative) constant parameter and $\sigma\,,\lambda$ positive numbers. As a consequence, the field behaves as
\begin{eqnarray}
\phi&\simeq&\phi_0-\Phi_0\left(\frac{2\kappa}{\kappa+\sigma-\lambda}\right)(N+1)^{\frac{\kappa+\sigma-\lambda}{2\kappa}}(-8\xi_0 H_0^2)^{1/(2\kappa)}\,,\quad -1<\lambda<0\,,
\nonumber\\
\phi&\simeq&\phi_0-\left(\frac{2\kappa}{\kappa-\sigma}\right)\Phi_0(1+N)^{\frac{\kappa-\sigma}{2\kappa}}(-8\xi_0H_0^2+1)^{1/(2\kappa)}\,,\quad 0=\lambda\,,\nonumber\\
\phi&\simeq&\phi_0-\left(\frac{2\kappa}{\kappa-\sigma}\right)\Phi_0(1+N)^{\frac{\kappa-\sigma}{2\kappa}}\,,\quad 0<\lambda\,,
\end{eqnarray}
with
\begin{equation}
\Phi_0=\left(\frac{2^{\kappa-1}}{\mu\kappa H_0^{2(\kappa-1)}}\right)^{1/(2\kappa)}\,.
\end{equation}
The potential reads
\begin{eqnarray}
V(\phi)&\simeq&3H_0^2\left[\Phi_0\left(\frac{2\kappa}{\kappa+\sigma-\lambda}\right)(-8\xi_0 H_0^2)^{1/(2\kappa)}\right]^{\frac{2\kappa}{\lambda-\sigma-\kappa}}(\phi_0-\phi)^{\frac{2\kappa}{\kappa-\lambda+\sigma}}
\,,\quad -1<\lambda<0\,,
\nonumber\\
V(\phi)&\simeq&
3H_0^2 \left[\left(\frac{2\kappa}{\kappa-\sigma}\right)\Phi_0(-8\xi_0H_0^2+1)^{1/(2\kappa)}\right]^{\frac{2\kappa}{\sigma-\kappa}}(\phi_0-\phi)^{\frac{2\kappa}{\kappa-\sigma}}
\,,\quad 0=\lambda\,,\nonumber\\
V(\phi)&\simeq& 3H_0^2 \left[\left(\frac{2\kappa}{\kappa-\sigma}\right)\Phi_0\right]^{\frac{2\kappa}{\sigma-\kappa}}(\phi-\phi_0)^{\frac{2\kappa}{\kappa-\sigma}}\,,\quad 0<\lambda\,,
\end{eqnarray}
while the coupling between $k$-essence and Gauss-Bonnet is given by
(\ref{27}).

Finally, the model (\ref{k2}) with (\ref{exxi}) and $g(\phi)$ as in (\ref{exxi2}) with $\sigma<\lambda$, reproduces the solution  (\ref{H2}) when the field behaves as
\begin{eqnarray}
\phi&\simeq&\phi_0-\Phi_0\left(\frac{2\kappa}{2\kappa-\lambda+\sigma}\right)(N+1)^{\frac{2\kappa-\lambda+\sigma}{2\kappa}}(-8\xi_0 H_{\text{dS}}^2)^{1/2\kappa}\,,\quad 0<\lambda<2\,,
\nonumber\\
\phi&\simeq&\phi_0-\Phi_0\left(\frac{2\kappa}{2\kappa-2+\sigma}\right)(-8\xi_0 H_\text{dS}^2+2)^{1/(2\kappa)}(N+1)^{\frac{2\kappa-2+\sigma}{2\kappa}}\,,\quad 2=\lambda\,,\nonumber\\
\phi&\simeq&\phi_0-\Phi_0\left(\frac{2\kappa}{2\kappa-2+\sigma}\right)2^{1/(2\kappa)}(N+1)^{\frac{2\kappa-2+\sigma}{2\kappa}}\,,\quad 2<\lambda\,,
\end{eqnarray}
with
\begin{equation}
\Phi_0=\left(\frac{2^{\kappa-1}}{\mu\kappa H_\text{dS}^{2(\kappa-1)}}\right)^\frac{1}{2\kappa}\,.
\end{equation}
The potential is reconstructed as
\begin{eqnarray}
V(\phi)&=&3H_0^2\left[1-2\left(\frac{2\kappa\Phi_0}{(\lambda-\sigma-2\kappa)(\phi_0-\phi)}(-8\xi_0 H_\text{dS}^2)^{1/(2\kappa)}\right)^\frac{2\kappa}{(2\kappa-\lambda+\sigma)}\right]
\,,\quad 0<\lambda<2\,,
\nonumber\\
V(\phi)&=&
3H_\text{dS}^2\left[1-2
\left(\frac{2\kappa\Phi_0}{(2\kappa-2+\sigma)(\phi_0-\phi)}(-8\xi_0 H_\text{dS}^2+2)^{1/(2\kappa)}\right)^\frac{2\kappa}{(2\kappa-2+\sigma)}
\right]
\,,\quad 2=\lambda\,,\nonumber\\
V(\phi)&=& 
3H_\text{dS}^2\left[1-2
\left(\frac{2\kappa\Phi_0}{(2\kappa-2+\sigma)(\phi_0-\phi)}(2)^{1/(2\kappa)}\right)^\frac{2\kappa}{(2\kappa-2+\sigma)}
\right]
\,,\quad 2<\lambda\,,
\end{eqnarray}
and the minimal coupling between $k$-essence and Gauss-Bonnet is given by (\ref{34}).

\section{Cosmological perturbations\label{pert}}

The cosmological perturbations around FRW metric (\ref{metric}) assume the form~\cite{Def, DeFelice},
\begin{equation}
ds^2=-[(1+\alpha(t, {\bf x}))^2-a(t)^{-2}\text{e}^{-2\zeta(t, {\bf x})}(\partial \psi(t,{\bf x}))^2]dt^2+2\partial_i\psi
(t,{\bf x})dt dx^i+a(t)^2
\text{e}^{2\zeta(t, {\bf x})}d{\bf x}\,,
\end{equation}
where $\alpha(t, {\bf x})\,,\psi(t, {\bf x})$ and $\zeta\equiv\zeta(t,{\bf x})$ are functions of the space-time coordinates. The action (\ref{action}) can be rewritten as
\begin{equation}
I=\int_\mathcal{M}dx^4 a^3 Q\left[\dot\zeta^2-\frac{c_s^2}{a^2}(\nabla\zeta)^2\right]\,,,\label{pertaction}
\end{equation}
where 
\begin{eqnarray}
&&
Q =
\frac{\phi'^2}{2H^2}
\left(
96H^4\xi'^2(\phi)+p_X(\phi, X)+\phi'^2 p_{XX}(\phi, X)
\right)\,,
\end{eqnarray}
and the
square of the sound speed reads
\begin{equation}
\hspace{-2cm}
c_s^2=
\frac{p_X(\phi, X)}{p_X(\phi, X)+96H^4\xi_\phi^2(\phi)+2p_{XX}(\phi, X)X}
\,.\label{c2}
\end{equation}
In this expressions, we have used the slow-roll approximation in the equations (\ref{eqslowroll}). We immediately see that in GR with classical canonical scalar field $c_s^2=1$, but here, thanks to non-GR effects and in the presence of $k$
-essence with $0<p_{XX}$, the square of the sound speed is smaller than one. By defining
\begin{equation}
v\equiv v(t, {\bf x})=z(t) \zeta(t, {\bf x})\,,\quad z\equiv z(t)=\sqrt{a^3 Q}\,,
\end{equation}
from (\ref{pertaction}) we infer
\begin{equation}
\ddot v_k+\left(k^2\frac{c_s^2}{a^2}-\frac{\ddot z}{z}\right)v_k=0\,,
\end{equation}
where we decomposed $v(t, {\bf x})$ in Fourier modes $v_k\equiv v_k(t)$ whose explicit dependence on $\bf k$ is given by $\exp[i {\bf k}{\bf x}]$. 
The solution for perturbations in quasi de-Sitter space-time can be derived as~\cite{DeFelice},
\begin{equation}
v_k(t)\simeq c_0\sqrt{\frac{a}{2}}\frac{a H}{(c_s k)^{3/2}}
\text{e}^{\pm i k\int \frac{c_s}{a}dt}
\left(1+i c_s k\int\frac{dt}{a}\right)\,,
\end{equation}
where the constant $c_0$ is fixed by the  Bunch-Davies vacuum state 
$v_k(t)=\sqrt{a}\exp[\pm i k\int c_s dt/a]/(2\sqrt{c_s\kappa})$ in the asymptotic past such that $c_0=i/\sqrt{2}$. Finally,
\begin{equation}
\zeta_k\equiv \frac{v_k}{\sqrt{A a^3}}=i\frac{H}{2\sqrt{A}(c_s k)^{3/2}}
\text{e}^{\pm i k\int \frac{c_s}{a}dt}
\left(1+i c_s k\int\frac{dt}{a}\right)\,,
\end{equation}
and the variance of the power spectrum of perturbations on the sound horizon crossing $c_s\kappa\simeq H a$ is given by
\begin{equation}
\mathcal P_{\mathcal R}\equiv\frac{|\zeta_k|^2 k^3}{2\pi^2}|_{c_s k\simeq H a}=\frac{H^2}{8\pi^2 c_s^3 A}|_{c_s k\simeq H a}\,.
\end{equation}
The spectral index is evaluated as
\begin{equation}
1-n_s=-\frac{d\ln \mathcal P_{\mathcal R}}{d \ln k}|_{k=a H/c_s}=2\epsilon+\eta_{sF}+s\,,
\end{equation}
where~\cite{DeFelice},
\begin{equation}
\epsilon=-\frac{\dot H}{H^2}\,,\quad \eta_{sF}=\frac{\dot \epsilon_s F+\epsilon_s\dot F}{H (\epsilon_s F)}\,,\quad s=\frac{\dot c_s}{H c_s}\,,\quad \epsilon_s=\frac{Q c_s^2}{F}\,,
\end{equation}
and
\begin{equation}
F=1+8(2-\log[X])X\xi_{\phi\phi}(\phi)\,.
\end{equation}
In a similar way, the tensor perturbations in flat FRW space-time leads to the tensor-to-scalar ratio,
\begin{equation}
r=16 c_s\epsilon_s\,.\label{rexp}
\end{equation}
In terms of the $e$-folds (\ref{N}), by taking into account (\ref{eqslowroll}), we derive
\begin{eqnarray}
(1-n_s)&\simeq&
\left(\phi ' \left(3072 H^6 p_X \xi '^3+24 H^4 \xi
   ' \left(p_X \left(16 p_X \xi ' \phi '^2+8
   \xi ''+p_{XX}\phi '^4\right)-12 \xi '
   p_X'\right)
\right.\right.
\nonumber\\&&
+H^2 \left(16 p_X^2 \xi ' \phi
   '^2+3 p_X^2 p_{XX} \phi
   '^6+\phi '^4 \left(p_X p_{XX}'-3
   p_{XX}p_X'\right)\right)
\nonumber\\&&
\left.\left.
+2 p_X^3 \phi
   '^4-2 p_X p_X' \phi '^2\right)-2
   p_X \phi '' \left(288 H^4 \xi '^2+H^2 p_{XX}
   \phi '^4+2 p_X \phi '^2\right)\right)
\nonumber\\&&
\times\frac{1}{2
   p_X \phi ' \left(96 H^4 \xi '^2+H^2 p_{XX}
   \phi '^4+p_X \phi '^2\right)}\,,
\end{eqnarray}
and
\begin{equation}
r\simeq
\frac{8 p_X\phi '^2
   \sqrt{\frac{p_X}{\frac{96 H^4 \xi
   '^2}{\phi '^2}+H^2 p_{XX} \phi
   '^2+p_X}}}{\frac{4 H^2 \left(2-\log \left(
   H^2 \phi '^2/2\right)\right) \left(\xi '' \phi '-\xi '
   \phi ''\right)}{\phi '}+1}\,.
\end{equation}
For canonical scalar field in the background of GR we recover the classical expressions:
\begin{equation}
(1-n_s)\simeq\phi'^2-\frac{2\phi''}{\phi'}\equiv
\frac{3H'}{H}-\frac{H''}{H'}
\,,\quad r\simeq 8\phi'^2\equiv\frac{16 H'}{H}\,.
\end{equation}
The last Planck satellite data~\cite{Planck} reveal
$n_{\mathrm{s}} = 0.968 \pm 0.006\, (68\%\,\mathrm{CL})$ and 
$r < 0.11\, (95\%\,\mathrm{CL})$. We see from (\ref{rexp}) that in the case of
$k$-essence models minimally coupled with Gauss-Bonnet, since $c_s<1$, the tensor-to-scalar ratio can be easily suppressed according with observations. On the other hand, by posing $N\equiv \mathcal N\simeq 60$ (it corresponds to the total e-folds necessary to have the thermalization of our universe), it is required that $(1-n_s)\simeq 2/N$ to have a viable scenario.

In the case of canonical scalar model in (\ref{dilaton}) with potential (\ref{potential1}) and coupling (\ref{27}), the (quasi)-de Sitter solution (\ref{H1}) leads to the spectral index
\begin{eqnarray}
(1-n_s)&\simeq&\frac{1+\lambda}{(1+N)}\,,\quad -1<\lambda<0\,,\nonumber\\
(1-n_s)&\simeq&\frac{2}{(1+N)}\,,\quad 0\leq\lambda\,.
\end{eqnarray}
The last case corresponds to chaotic inflation in GR with quadratic potential and the spectral index is in agreement with Planck data when $N\simeq 60$. We note that even if the Gauss-Bonnet contributes to the solution (this is the case $\lambda=0$), it does not enter in the spectral index. However, in the case of pure canonical scalar field with quadratic potential ($0<\lambda$, when the coupling between field and Gauss-Bonnet is negligible), a simple calculation leads to a tensor-to-scalar ratio slightly larger than observations, namely $r\simeq 8/(N+1)\simeq 0.13$. Thus, if we take $0=\lambda$, we may introduce a small correction in the tensor-to-scalar ratio through the Gauss-Bonnet reducing the sound speed in (\ref{c2}) and recovering the correct observed value. When $-1<\lambda<0$, the contribution of Gauss-Bonnet significatively changes the dynamics of inflation with respect to the classical case and the scenario is not viable.

For the inflationary solution (\ref{H2}) with potential (\ref{33}) and coupling (\ref{34}), we have
\begin{eqnarray}
(1-n_s)&\simeq&\frac{\lambda}{(1+N)}\,,\quad 0<\lambda<2\,,\nonumber\\
(1-n_s)&\simeq&\frac{2}{(1+N)}\,,\quad 0\leq \lambda\,.
\end{eqnarray}
We can see that for $0\leq\lambda$, the Gauss-Bonnet does not contribute to the spectral index and inflation is viable (it corresponds to a Starobinsky-like inflation in the Einstein's frame~\cite{Staro}). On the other hand, for $0<\lambda<2$, the perturbations depend on the coupling between the field and the Gauss-Bonnet, and we find the Planck data only if $\lambda$ remains close to two.

For the $k$-essence models considered in subsection \S~\ref{sub32}, we can verify that when the Gauss-Bonnet does not significatively enter in the dynamics of inflation (namely, when $0\leq \lambda$ and $2\leq\lambda$ of solutions (\ref{H1}) and (\ref{H2}), respectively), the spectral index assumes the same (viable) form of the canonical scalar field models, such that $(1-n_s)\simeq 2/(N+1)$. 
The similarity between canonical fields and fields with higher order kinetic term seems to remain also in the presence of the Gauss-Bonnet contribution.
For example, in the case of solution (\ref{H2}), when $0<\lambda<2$ and  independently on the values of $\kappa$ and $\sigma$, in both of the models (\ref{k12}) and (\ref{k2}) one has $\lambda/(1+N)$.

\section{Conclusions}

In this paper, we investigated solutions for inflation in a class of Horndeski models where the scalar field representing inflaton is coupled with the Gauss-Bonnet invariant. We considered different scenarios, namely canonical scalar field with standard kinetic term and $k$-essence, where scalar field with higher order kinetic term is present. The interest in this kind of theories in the context of early-time inflation is motivated by the fact that one expects that at high curvatures non-GR effects modify the theory of Einstein. In the specific, the Gauss-Bonnet invariant plays an important role in string theory and enters in the trace anomaly (see also Ref.~\cite{buch,quattro,cinque,sei,sette}). Moreover, $k$-essence is one of the possible descriptions for inflation (see also Refs.~\cite{ultimok, ultimok2}). The advantage for dealing with a Horndeski model is that, despite the involved form of the Lagrangian, the equations of motion are at the second order like in General Relativity.

Different behaviors of the Hubble parameter for early-time acceleration have been analyzed, by making use of a reconstruction  technique to infer Lagrangians by starting from the given solutions. 
This procedure is quite simple if one expresses all the quantities in terms of the $e$-folds number which measures the amount of inflation. In the last section, we also
furnished the formalism to calculate perturbations in our class of models deriving the spectral index and the tensor-to-scalar ratio. We applied and discussed the results to our examples (for a comparison, see also Ref.~\cite{RGinfl}). 

One can further generalize our action where the field is linearly coupled with the Gauss-Bonnet  invariant by considering the coupling of arbitrary Gauss-Bonnet modified gravity with $k$-essence, in analogy with similar $F(R)$-$k$-essence models analyzed in Ref.~\cite{OdkFR}.
For some other reviews of inflation in modified gravity see Refs.~\cite{Odinfrev, myrevinfl}.  Other useful references can be found in Refs.~\cite{uno, unobis, due, tre}.


\end{document}